# Anticoincidence Detector for GLAST

**Alexander Moiseev, Jay Norris, Jonathan Ormes, Steven Ritz, and David Thompson**
on behalf of the GLAST collaboration
*NASA/Goddard Space Flight Center, Greenbelt, MD 20771, USA*

**Abstract**
The Gamma-ray Large Area Space Telescope (GLAST) is now being designed by a number of collaborating institutions. It will study the cosmic gamma radiation from 20 MeV to 300 GeV with high precision and sensitivity, greatly expanding on the important EGRET results. One of the key systems of the instrument, the Anticoincidence Detector (ACD), is designed to reject the majority of charged particles, which are the background for any gamma-ray experiment. The ACD of EGRET has suffered from the self-veto effect when the products of the high energy photon interactions in the instrument's calorimeter cause a veto signal in the anticoincidence detector (backsplash effect), resulting in the degradation of the efficiency for high energy (> 5 GeV) gamma rays. To avoid this effect, the ACD for GLAST is divided into many scintillating tiles with wave-shifting fiber readout. The design of this detector along with the beam test and simulation results are given in this paper.

## 1. Introduction

In space, a high energy gamma ray telescope must identify gamma rays against an enormous background of charged particle cosmic rays that bombard the instrument from all directions. GLAST (Atwood et al., 1994) science depends on being able to discriminate with high accuracy against this charged particle background. The scientific requirement that drives the instrumental background rejection is the study of the diffuse high latitude gamma-radiation.

A plastic scintillator anti-coincidence detector (ACD) provides a high-reliability, high-efficiency, low-cost approach to screening out the cosmic ray background. Similar systems have been used for a number of gamma-ray telescopes in past: SAS-2, COS-B and EGRET/CGRO, all of which used a monolithic plastic scintillator viewed by photomultiplier tubes (PMT). GLAST, in addition to having much higher sensitivity at low energies, will extend the covered energy span up to 300 GeV to overlap significantly with ground-based instruments. The highest energy gamma-rays, especially with energies above 5 GeV, produce backsplash – mainly low-energy, minimum-attenuation photons, originating in the calorimeter as the products of the electromagnetic shower. Such backsplash photons can cause a veto pulse in the ACD through Compton scattering. The EGRET telescope suffered a ~50% loss of efficiency at 10 GeV relative to 1 GeV due to this effect ( Thompson et al., 1993). This self-veto is reduced by segmenting the GLAST ACD into tiles and vetoing an event only if the pulse appears in the tile through which the reconstructed event trajectory passes. The segmentation optimization is one of the issues being studied in this work.

## 2. Requirements for the ACD
**2.1. Background rejection:** To be able to analyze the high-latitude diffuse gamma-radiation, we require that the background due to charged particles be less than 10% of the diffuse signal. For setting background rejection criteria, we care about both integral and differential fluxes of protons and electrons (Ormes, 1999, and references therein). The most problematic decade is that between 3 and 30 GeV. At lower energy, at least outside the radiation belts, the geomagnetic cutoff reduces the cosmic ray rates dramatically. Since we are not going to operate GLAST inside the belts, the very soft and very intense charged particles fluxes there will not be of concern. At higher energies, the cosmic ray fluxes fall faster than the high latitude diffuse gamma-ray flux.

The maximum integral proton flux at solar minimum is $2 \times 10^5$ times the diffuse high latitude photon flux (above 3 GeV) at the lowest vertical geomagnetic cutoff in the planned 28 degree orbit. To set the requirements for the ACD we consider this worst case, although this condition applies to only a small part of the orbit. This worst case is important in estimating the second level trigger rate, specifically the proton rejection needed at this level in order to manage the data rate. The problem of proton contamination is less severe because the efficiency for the calorimeter to capture the energy of the proton is low. The ratio of protons that deposit energy > 3 GeV in the calorimeter to the diffuse high latitude photons of 3 GeV and above is about $6 \times 10^3$. These protons can be rejected by using all the GLAST detectors – looking at the topologies of the track in the tracker and of the shower in the calorimeter as well as the signal in the ACD. In particular, the proton-induced shower in a calorimeter very rarely looks like an electromagnetic one caused by a photon. The cosmic-ray electron flux is the more serious background though it is ~100 times less intense than the protons, because the calorimeter has no power to distinguish electrons from photons, and the ACD is the main defense. The ratio of cosmic electrons to high latitude photons peaks at 3000 between 3 and 10 GeV.

Based on the above considerations, we set the GLAST system level design requirement for the misidentification of charged cosmic rays to be $3 \times 10^{-5}$ or a detection efficiency of 0.99997, from which three orders of magnitude rejection will be provided by the tracker+calorimeter for protons and one order of magnitude for electrons. The electron component then drives **the requirement for the ACD to have 0.9997 efficiency and hermeticity for singly-charged particles**.

**2.2. Backsplash avoidance:** We set the goal to have a photon detection efficiency degradation due to backsplash to be less than 10% over the entire GLAST energy range up to 300 GeV.

**2.3. Flight constraints:**
- mass less than 150 kg
- electrical power less than 30 Watts
- dimensions 170 cm × 170 cm × 60 cm to cover the tracker from the top with not more than 5 cm total thickness on the sides
- minimal inert material outside the ACD to prevent additional instrumental background due to proton-induced gamma-ray production
- robust to launch loads

## 3. Beam test

A prototype of the ACD was tested at SLAC in the fall of 1997 (Atwood et al., 1999). Wave-shifting fiber (WSF) readout was chosen as providing a uniform detection efficiency over the detector area. The goals of the experiment were to test the readout technique and to study carefully the backsplash effect for the ACD segmentation optimization. The beam test

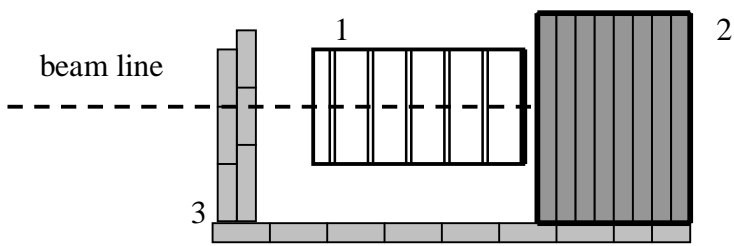

Fig.1. Beam test set up. 1 – tracker, 2 – CsI calorimeter, 3 – ACD scintillator paddles

ACD consisted of two modules, as shown in fig.1. One module contained 9 scintillator paddles (made of BC-408) placed on the side of the tracker/calorimeter, and the other module, placed just upstream of the tracker, consisted of two superimposed layers with 3 paddles in each. WSF were embedded in grooves across the face of the 1-cm thick paddles to transfer the light to Hamamatsu R-647 phototubes. The signal from each phototube was pulse-height analyzed by a CAMAC 2249A PHA module along with the beam trigger and information from the tracker and calorimeter. Use of the paddles provides a knowledge of the backsplash angular distribution, which is shown in fig.2a for 25 GeV photons. The efficiency for singly-charged particle detection was measured for 25 GeV electrons in this beam test and using cosmic ray muons in the laboratory, and shown in fig.2b. The required efficiency of 0.9997 is achieved at a threshold of 0.2 of a minimum ionizing particle energy deposition.

The obtained results were used to verify Monte Carlo simulations, which were then extended to 300 GeV to optimize the GLAST ACD segmentation. Simulations were done with the event generator

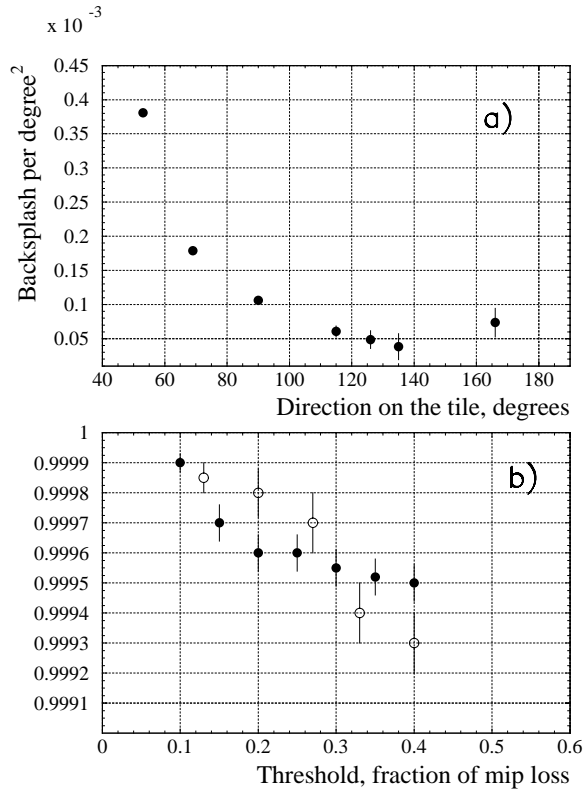

Fig.2. a) – Backsplash angular distribution
  b) – efficiency of scintillator paddle with WSF readout, filled circles – measured in a beam, opened circles – measured with cosmic ray muons the tracker but have a reasonable path in the calorimeter. For the baseline design the segmentation on the sides is of about the same size as on the top. In order to optimize the search for high-energy gamma-ray lines originating in dark matter annihilation, the segmentation on the sides should probably be decreased to ~ 200 cm$^2$ (Moiseev et al., 1999), and our intention is to design such a system within the ACD mass and power constraints.

## References

Atwood, W. B., et al. 1994, NIM A342, 302
Atwood, W.B., et al. 1999, submitted to NIM
Moiseev, A.A., et al. these Proccedings
Ormes, J.F. 1999, GSFC/LHEA internal report
Thompson, D.J., et al. 1993, ApJ 86, 629

Glastsim. Results of these simulations indicate that a tile size of ~1000 cm$^2$ will be sufficient for the top surface of the ACD to maintain 90% efficiency against self-veto by backsplash up to 300 GeV (fig.3).

## 4. Summary – flight unit design approach

Plastic scintillator is a logical choice for an anticoincidence detector because it is rugged, space-qualified, low cost, highly efficient for charged particle detection and easy to work with. The flight ACD consists of scintillator tiles, each 1 cm thick and ~1000 cm$^2$ of area. Each tile is read out by wave-shifting fibers (BCF-91A) to provide the uniform sensitivity over the area and to avoid placing phototubes, which are additional material, in the instrument aperture. The spacing between fibers is about 1 cm, and they are connected to two remote phototubes for redundancy. Tiles are overlapped in one direction and glued on the other to provide the best hermeticity. The ACD "hat" covers the top and the sides of the tracker down to the calorimeter – for the most effective use of the GLAST aperture, making it sensitive to the events which miss most of

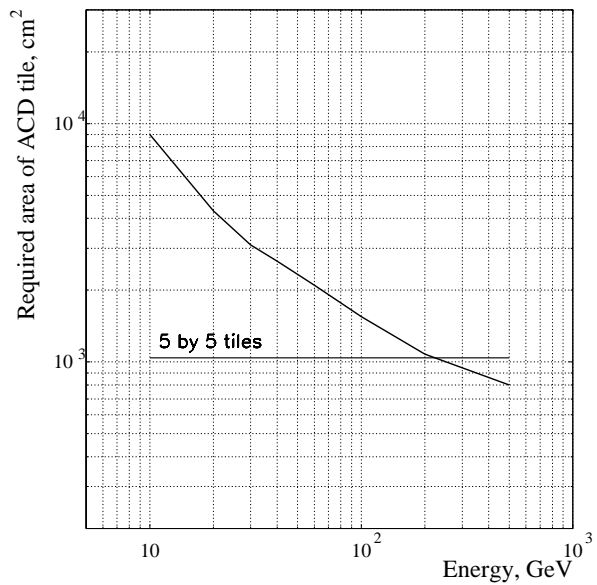

Fig.3. Area of ACD tile required to maintain 90% efficiency – extrapolation of beam test results by Glastsim simulations